\documentclass[11pt,twocolumn,aps,prapplied,reprint,superscriptaddress ] {revtex4-1}

\usepackage{natbib}

\usepackage{pst-node,graphicx}
\usepackage{braket}
    
\usepackage{graphicx,amsmath,amssymb,braket}

\usepackage[colorlinks=true,linkcolor=blue]{hyperref}
\usepackage{ulem}

\begin{document}

\title{Signatures of non-local conductivity in near-field microscopy
}

\author{Mikhail Khavronin}
\affiliation{Laboratory of 2d Materials for Optoelectronics, Center for Photonics and 2d Materials, Moscow Institute of Physics and Technology, Dolgoprudny 141700, Russia}

\author{Dmitry Svintsov}
\affiliation{Laboratory of 2d Materials for Optoelectronics, Center for Photonics and 2d Materials, Moscow Institute of Physics and Technology, Dolgoprudny 141700, Russia}

\begin{abstract}
We propose and theoretically substantiate a new method to study the non-local conductivity of two-dimensional electron systems (2DES) using the tools of near-field microscopy. We show that the height dependence of induced dipole moment of illuminated near-field probe is substantially different for various transport regimes of charge carriers in 2DES. For hydrodynamic transport regime, the induced dipole moment scales as $z_0^{-2}$, where $z_0$ is the elevation of probe above the 2DES. Both for Drude and classical ballistic regimes of conduction, the dipole moment scales as $z_0^{-3}$. In the former case, the dipole moment is carrier density-independent, while in the latter it largely depends on carrier density. More generally, we find that the induced dipole moment of the probe is proportional to the Laplace transform of wave-vector dependent conductivity and inverse dielectric function of 2DES over the wave vectors $q$. Our results should provide a simple tool for studies of non-local conductivity in solids that was challenging to address with other techniques. 
\end{abstract}

\maketitle

\section{Introduction}
The relations between electric field and current in solids are generally non-local, which implies that current at a given point $\mathbf{j}(\mathbf{r})$ can be affected by field at remote positions ${\bf E}(\mathbf{r'})$:
\begin{equation}
    \mathbf{j}(\mathbf{r}) = \int \sigma(\mathbf{r},\mathbf{r'}){\bf E}(\mathbf{r'}) d\mathbf{r'}
\end{equation}
The non-locality comes from thermal or quantum motion of charge carriers. Formally, this motion results in explicit dependence of conductivity kernel $\sigma({\bf r},{\bf r}')$ on positions ${\bf r}$ and ${\bf r}'$. At zero frequency, the conductivity kernel is large at distances $|{\bf r}' - {\bf r}|$ less than electron mean free path. At finite frequencies, the kernel typically decays at electron path during the field cycle $l_\omega = v_0 / \omega$, where $v_0$ is the thermal or Fermi velocity of charge carriers~\cite{Physical_Kinetics}. At even smaller (quantum) distances $|{\bf r}' - {\bf r}|$, the conductivity kernel may possess extra features associated with Friedel oscillations or cyclotron motion~\cite{Chiu_plasma_oscillations} of electrons in magnetic field.

Once the non-local conductivity kernel $\sigma(\mathbf{r},\mathbf{r'})$ is known, it may be tempting to decode the information about carrier dynamics from it~\cite{Basov_RMP_GrapheneSpectroscopy}. Such method may become a simple complement to complex angle-resolved photoelectron spectroscopy. Most straightforward ways to measure the non-local dielectric function rely on electron energy loss spectroscopy~\cite{EELS-nonlocal}, which requires ultra-high vacuum and atomically clean surfaces. There exist all-electrical methods for studies of non-local conduction, where current is injected between a couple of contacts, while the voltage is measured between another couple~\cite{Non-local-1,Non-local-2,Non-local-3,Non-local-4}. Such technique, however, does not enable a continuous measurement of $\sigma$ as a function of distance; instead, it is limited by an initially defined contacts' geometry. Recently, indirect evidence of transport non-locality were theoretically revealed in height-dependent magnetic noise above 2DES~\cite{Agarwal_noise}.  

In recent years, a great attention is attracted to the technique of near-field optical microscopy by scattering from the tip~\cite{sSNOM-review}. It enables the reconstruction of optical properties in non-uniform structures [dielectric function $\varepsilon({\bf r})$ or surface conductivity $\sigma({\bf r})$] with resolution reaching $\sim 10^{-3}\lambda_0$~\cite{s-SNOM-resolution,s-SNOM-MOSFET,Govyadinov_permittivity_1,Govyadinov_permittivity_2}. All such studies, however, assume the local relations between current and electric field, ${\bf j}({\bf r}) \approx \sigma({\bf r}) {\bf E}({\bf r})$. Attempts to extract the non-local conductivity (even in uniform structures) from near-field measurements are yet on their initial stage. In particular, the non-locality of conductivity can affect the speed of collective excitations -- plasmons~\cite{Ryzhii_plasmons,DasSarma_plasmons}, which, in turn, can be extracted from polariton interferometry~\cite{Lundeberg_quantum_nonlocal,Non-locality-Kane-Plasmons}. Such technique allows one to extract the information about non-locality only at 'interlocked' values of frequency $\omega$ and wave vector $q$ satisfying the plasmon dispersion relation $\varepsilon(q,\omega) = 0$.

\begin{figure}[ht!]
\center{\includegraphics[width=0.9\linewidth]{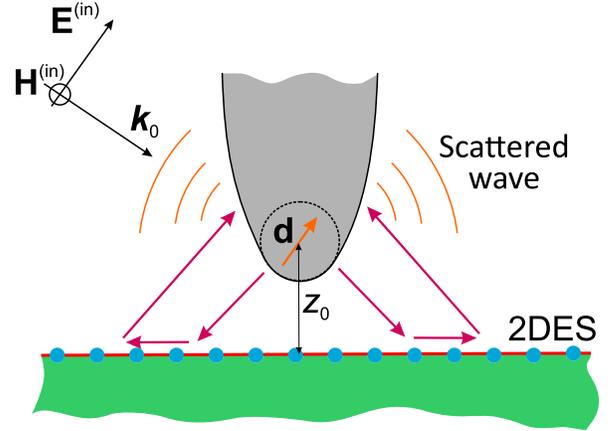}}
\caption{Schematic of the setup. A near-field probe (grey) is illuminated by incident plane wave. Its near field (purple) is reflected by a two-dimensional electron system (2DES), which reacts to the field non-locally. The reflected wave modifies the dipole moment of the probe ${\bf d}$. This dipole produces scattered far-field which carries information about the surface conductivity of 2DES}
\label{fig:Scheme}
\end{figure}

In this paper, we substantiate theoretically a new method for studies of non-local carrier dynamics from the near-field optical signals. We show that electromagnetic scattering from a near-field probe located at small distance $z_0$ above the 2d conductor is strongly affected by the non-locality of conductivity kernel. More precisely, the small-height asymptotic of induced dipole moment of the probe ${\bf d}(z_0) - {\bf d}(\infty) = \alpha_{\rm eff}{\bf E}$ is a power-law function $\propto z_0^{-n}$. The exponent of this dependence $n$ is linked to the asymptotics of Fourier-transformed non-local conductivity $\sigma({\bf q},\omega)$ at large field momenta ${\bf q}$. As a prototypical example, we consider the scaling of near-field signals for 2d electrons obeying the laws of drift and diffusion (Drude conduction), exhibit a classical ballistic motion, or obey the laws of hydrodynamics. We find that $\alpha_{\rm eff} \propto z_0^{-3}$ in the two former cases, and $\alpha_{\rm eff} \propto z_0^{-2}$ in the case of hydrodynamics. More generally, we find that induced dipole moment $d_z(z_0)$ of the near-field probe can be presented as a Laplace transform of non-local conductivity $\sigma(q,\omega)$ divided by 2d dielectric function $\varepsilon_{2D}(q,\omega)$ with respect to the field momentum $q$.

\section{Solution of scattering problem for non-local conductivity}

 The studied system represents an extended two-dimensional electron system (2DES) located at $z=0$ and a near-field probe located at ${\bf r}_\parallel =0 $ elevated at height $z_0$ (Fig.~\ref{fig:Scheme}). The probe is illuminated by a plane wave which polarises the probe and generates electromagnetic near fields with large Fourier harmonics ${\bf E}_{{\bf q},\omega}$. These near fields are reflected from 2DES; their refection coefficient is generally determined by non-local conductivity $\sigma({\bf q},\omega)$. Reflected waves modify the dipole moment of the probe ${\bf d}$ and therefore modify its far field radiation. One may suggest that the characteristic wave vector or tip-induced near fields is $q \lesssim z^{-1}_0$. By placing the tip at progressively smaller distances, one collects information about surface conductivity at larger and larger wave vectors.
 
To justify the suggested scheme, we present an exact solution for the scattering problem presented in Fig.~\ref{fig:Scheme} with full account for conduction non-locality. For analytical traceability, the tip is modelled as a point dipole with moment ${\bf d} e^{-\omega t}$. At the first stage, we find the fields in all space ${\bf E}^{(d)}({\bf r})$ provided that ${\bf d}$ is fixed, and assuming no external illumination. A similar problem of dipole radiation above the non-local surface is known in for bulk metals~\cite{Dipole_non_local_metal}, but has not been reported for 2D. This is done, most conveniently, by solving the wave equation for the vector potential in the Lorentz gauge:
\begin{equation}
\left(k_z^2-\frac{d^2}{dz^2}\right)\mathbf{A}^{(d)}_{\omega q}(z)  =\frac{4\pi}{c} \left[ \mathbf{j}_{\omega k}^{(d)} + \mathbf{j}_{\omega q}^{(2des)}\right].
\end{equation}
In the above equation, we have introduced the Fourier transform with respect to the in-plane coordinate, ${\bf A}_{q\omega} = \int{A_\omega({\bf r}_\parallel)}e^{i {\bf q}{\bf r}_\parallel} d{\bf r}_\parallel$, $k_z^2 = q^2 - (\omega/c)^2$ is the squared transverse wave vector, $\mathbf{j}_{\omega q}^{(d)} = -i \omega {\bf d} \delta(z - z_0)$ is the current density at the oscillating dipole, and $\mathbf{j}_{\omega q}^{(2des)}$ is the distribution of surface current density at the 2DES. We link it to the electric field via non-local Ohm's law:
\begin{equation}
 \mathbf{j}_{q \omega}^{(2des)} = \sigma(q,\omega) {\bf E}^{(d)}_{q\omega}(z=0) \delta(z),
\end{equation}
while the in-plane electric field is obtained via vector-potential as
\begin{gather}
{\bf E}^{(d)}_{q\omega} = \frac{i \hat D}{\omega}{\bf A}^{(d)}_{q\omega} ,\\
\hat{D} = \left(
\begin{array}{cccc}
q_0^2-q_x^2 & -q_x q_y & i q_x \frac{\partial}{\partial z}\\
-q_x q_y &q_0^2- q_y^2 & i q_y \frac{\partial}{\partial z}\\
0 & 0 & 0
\end{array}
\right).
\end{gather}
Combining the above equations, we arrive at a simple second-order equation for the vector-potential with two delta-sources in the right-hand side:
\begin{multline}
\left(k_z^2-\frac{d^2}{dz^2}\right)\mathbf{A}^{(d)}_{\omega q}(z)  =\\
\frac{4\pi}{c} \left[ -i \omega {\bf d} \delta(z-z_0) + \frac{i  \sigma_{\omega q}}{\omega}  \delta(z)\hat{D}\mathbf{A}^{(d)}_{\omega q}(0) \right].
\end{multline}
The above equation is readily solved, and the full solution is a linear function of the dipole moment amplitude $\mathbf{A}^{(d)}_{\omega q}(z) \propto {\bf d}$.

At the second stage, we use the superposition principle to find the total field created by external illumination and fixed dipole:
\begin{equation}
 {\bf E}_\omega ({\bf r}) = {\bf E}^{(\rm in)}_\omega ({\bf r}) + {\bf E}^{(\rm r)}_\omega ({\bf r}) + {\bf E}^{(d)}_\omega ({\bf r}),
\end{equation}
where ${\bf E}^{(\rm in)}$ and ${\bf E}^{(\rm r)}$ are the fields of incident and reflected waves in the presence of 2DES but in the absence of the probe, and ${\bf E}^{(d)}$ is the field of the probe with fixed dipole moment ${\bf d}$.

At the last stage, we release the assumption of the fixed dipole moment and link it to the local field ${\bf E}_\omega ({\bf r}_0)$ via the polarizability $\alpha$:
\begin{multline}
\label{Self-consistency}
    {\bf d} = \alpha {\bf E}_\omega ({\bf r}_0) = \\
    \alpha\left[{\bf E}^{(\rm in)}_\omega ({\bf r_0}) + {\bf E}^{(\rm r)}_\omega ({\bf r_0}) + \int{\frac{d^2 {\bf q}}{(2\pi)^2} {\bf E}_{\omega q}^{(d)} (z_0) } \right].
\end{multline}
Certain care should be taken upon evaluation of the last integral. Strictly speaking, it diverges because the field created by point dipole at its own origin is infinite. This self-action term should be subtracted; as a result, the dipole is polarized according to the magnitude of smooth fields modified by the presence of 2DES. Solving Eq.~(\ref{Self-consistency}) which is linear with respect to ${\bf d}$, we find:
\begin{equation}
\label{Dip-moment}
    d_i = \alpha \frac{ {E^{(in)}_{i}}+{E^{(r)}_{i}}}{1-\alpha  I_i},
\end{equation}
where $i=\{x,y,z\}$ labels the coordinate axes, and $I_i$ is the polarization factor. Its expression is the simplest for incoming field polarized along the $z$-axis:
\begin{equation}
\label{I}
   I_z =\frac{1}{k_0} \int_0^{\infty}  \frac{ 2\pi i \sigma(q,\omega)}{c}\frac{q^3}{ \epsilon_{2D}(q,\omega)}  e^{-2 k_z z_0 } dq,
\end{equation}
where 
\begin{equation}
\epsilon_{2D}  (q,\omega) = 1 + \frac{2 \pi i  \sigma(q,\omega)}{\epsilon_b \omega /q}    
\end{equation}
is the effective dielectric permittivity of 2DES, and $\epsilon_b$ is the background dielectric permittivity. The result coincides with that reported in [\onlinecite{Fogler_quant_near_field}] if expressed through wave-vector dependent reflection coefficients.

Generally, the polarizability of the near-field probes is very small, order of $r^{3}_{\rm tip}$, where $r^{3}_{\rm tip} \sim 10$ nm is the curvature radius of the tip.  This justifies the expansion of Eq.~\ref{Dip-moment} in powers of $\alpha$:
\begin{equation}
     d_{i} = ({E^{in}_{i}}+{E^{r}_{i}})(\alpha + \alpha^2  I_{i} + \alpha^3  I_i^2+...).
\end{equation}
The linear-in-$\alpha$ term contains no information about near fields, it is sensitive only to the reflection of incident plane wave from uniform 2DES. The $\alpha^2$-term is the largest one that carries information about the near-field reflection. It will be in the focus of subsequent analysis, while the quantity $\alpha I_i$ will be called the effective polarizability.

\section{Analysis of induced dipole moment for particular transport regimes}
The dipole moment of the near-field probe contains information about non-local conductivity. Indeed, according to Eq.~(\ref{I}), $I_z$ is a convolution of wave-vector dependent conductivity $\sigma(q,\omega)$, the inverse permittivity $\epsilon^{-1}_{2D}(q,\omega)$, and the height-dependent factor $e^{-2\sqrt{q^2 - k_0^2}z_0}$. We shall further analyze the height dependence of effective polarizability and show that it indeed depends on the regime of carrier transport in 2DES.

Before proceeding, we note that in far-field zone, $z_0 \gg \lambda_0$, the effective polarizability displays an inverse proportionality to $z_0$.  This result is independent of particular transport regime in 2DES. Indeed, in the far zone, the wave vectors $q \leq k_0$ yield the dominant contribution to the integral (\ref{I}). The spatial dispersion of conductivity does not develop at such small wave vectors; it occurs only at $q \sim \omega/v_0 \gg \omega/c$. For this reason, all further considerations will be restricted to the near-field region $k_0z_0 \ll 1$.


\subsection{Drude conductivity}

\begin{figure}[ht]
\center{\includegraphics[width=0.9\linewidth]{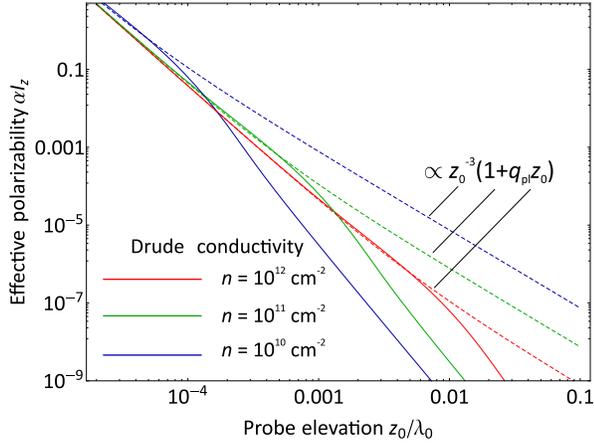}}
\caption{Effective polarizability of the near-field probe $\alpha I_z$ vs the probe elevation $z_0$ (in units of free space wavelength $\lambda_0$) for Drude conduction in 2DES. Solid lines are results of exact integration (Es.~\ref{I}), dashed lines are power-law asymptotes. Incident wave frequency $\omega/2\pi = 1$ THz, momentum relaxation time $\tau_p = 2$ ps, effective mass $m = 0.067m_0$, background dielectric constant $\epsilon_b = 4$, tip radius $r_{\rm tip} = 10$ nm}
\label{fig:Drude}
\end{figure}

The simplest model of 2DES conductivity is the Drude model, wherein the spatial dispersion is absent at all:
\begin{equation}
\label{Sigma_drude}
\sigma_D=\frac{n e^2}{m_e {(\omega+i/\tau_p)}},
\end{equation}
here $n$ is the density of 2d electrons, $m_e$ is their effective mass, and $\tau_p$ is the momentum relaxation time. The resulting height-dependent effective polarizability is shown in Fig.~\ref{fig:Drude} for three characteristic carrier densities $n = 10^{12}$ cm$^{-2}$, $10^{11}$ cm$^{-2}$ and $10^{10}$ cm$^{-2}$ and $r_{\rm tip} = 10$ nm. The overall dependence has two distinct regions with different height scalings.

The effective polarizability in this case is evaluated analytically to yield:
\begin{multline}
   I_z = \frac{i\pi \sigma_D }{c}\frac{q_{\rm pl}}{k_0}\times \\
   \left\{q_{\rm pl}^3 e^{-2 q_{\rm pl}z_0} (\text{Ei}(2 q_{\rm pl}z_0)-i \pi )-\frac{q_{\rm pl}z_0 (2 q_{\rm pl}z_0+1)+1}{4 z_0^3}\right\},
\end{multline}
where it was convenient to introduce the wave vector of 2d plasmons:
\begin{equation}
    q_{\rm pl}=\frac{\epsilon_b \omega (\omega + i/\tau_p)}{2 \pi n_e e^2/m_e}.
\end{equation}
The only dimensionless parameter governing the principal height dependence of $I_z$ is the ratio of probe elevation and 2d plasmon wavelength. At moderate heights, $q_{\rm pl}z_0 \gg 1$ (but still $z_0 k_0 \ll 1 $), the height dependence follows the $z_0^{-4}$ asymptotics:
\begin{equation}
    I_z(q_{\rm pl}z_0 \gg 1)\approx \frac{3i\pi}{8}\frac{\sigma_D}{c}\frac{1}{k_0z_0^4}.
\end{equation}
At even smaller distances, $q_{\rm pl}z_0 \ll 1$, the scaling of polarizability is inverse cubic:
\begin{equation}
    I_z(q_{\rm pl}z_0 \gg 1) \approx \frac{\epsilon_b}{8z_0^3}(1 + q_{\rm pl}z_0).
\end{equation}
It is remarkable that the leading term of induced dipole moment is independent of carrier density at small heights. Indeed, at large $q$, the dielectric function of 2DES $\epsilon_{2D}$ is directly proportional to conductivity. The integrand in (\ref{I}) becomes independent of surface conductivity; we may speculate that the 2D conductor acts as a perfect mirror in this limit.

\subsection{Hydrodynamic  transport}
\begin{figure}[ht]
\center{\includegraphics[width=0.9\linewidth]{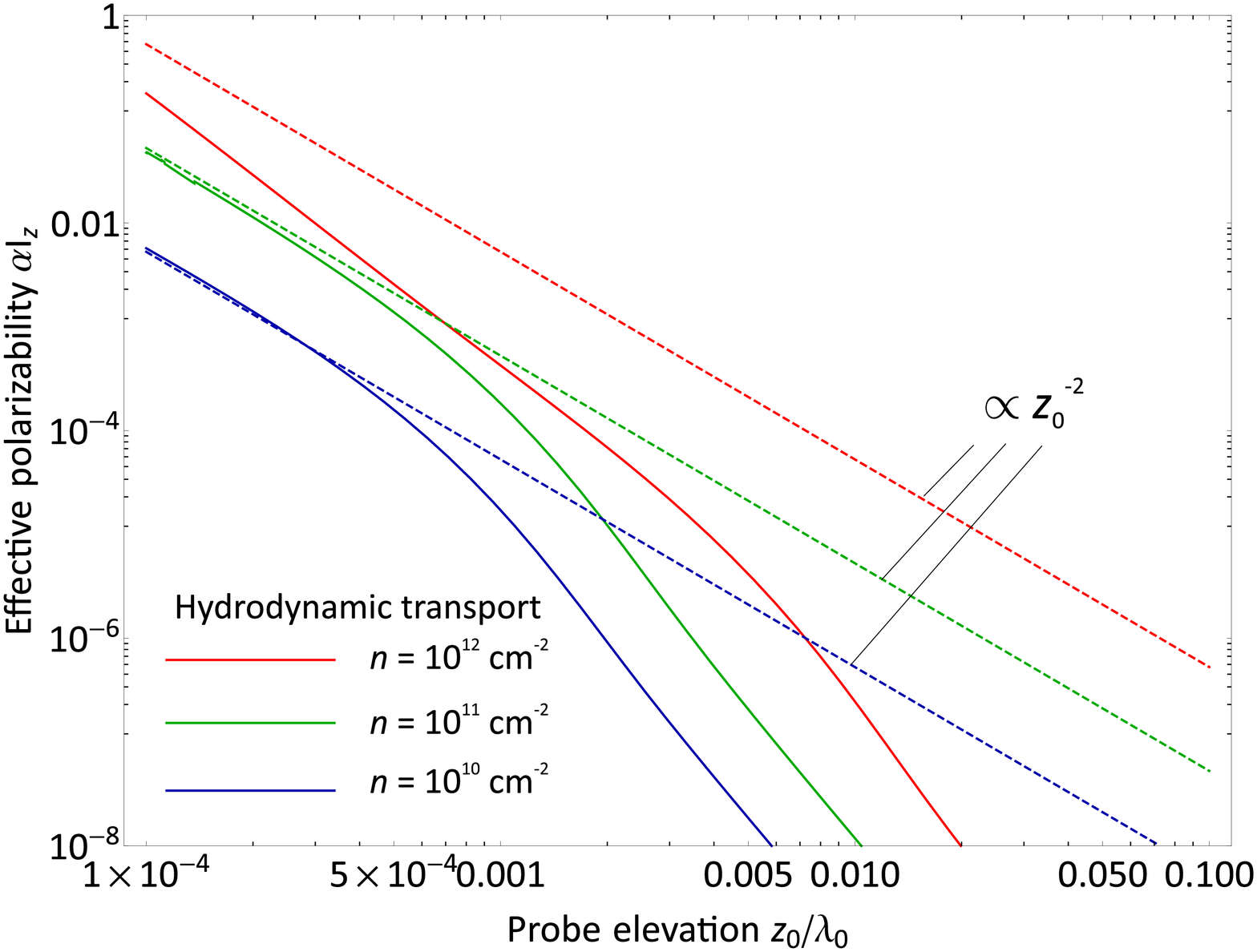}}
\caption{Effective polarizability of the near-field probe $\alpha I_z$ vs the probe elevation $z_0$ for hydrodynamic transport in 2DES. Solid lines are results of exact integration (Es.~\ref{I}), dashed lines are power-law asymptotes. Incident wave frequency $\omega/2\pi = 2$ THz, momentum relaxation time $\tau_p = 2$ ps, effective mass $m = 0.067m_0$, background dielectric constant $\epsilon_b = 4$, tip radius $r_{\rm tip} = 10$ nm, Fermi velocity $v_0 = 10^6$ m/s}
\label{fig:Hydro}
\end{figure}

The hydrodynamic transport mode in 2DES is established if the carrier-carrier collisions are so rapid that electrons behave as a viscous fluid. Parametrically, this corresponds to $\omega \tau_{ee} \ll 1$ and $q l_{\rm fp} \ll 1$, where $\tau_{ee}$ is the mean free time between electron-electron collisions and $l_{\rm fp} = v_0 \tau_{ee}$ is the mean free path~\cite{Abrikosov_Fermi_liquid,Svintsov_crossover}. In this limit, the conductivity is given by:
\begin{equation}
\label{Sigma_hd}
\sigma_{\rm hd}=\frac{n e^2 \omega/m_e}{\omega (\omega+i/\tau_p)-v_0^2 q^2/2}.    
\end{equation}
The full plot of polarization factor $I_z$ in such transport mode is shown in Fig.~\ref{fig:Hydro}. At very low heights, $z_0 \lesssim v_0/\omega$, the scaling is inverse quadratic (compared to inverse cubic for Drude conductivity). The reason for difference lies in asymptotic behaviour of conductivity $\sigma_{\rm hd} \propto q^{-2}$ at large wave vectors. The conductivity decays very rapidly at large wave vectors (small heights), thus the 2DES does not actively reflect the electromagnetic near fields. As a result, it fails to build up large dipole moment of the probe. This contrasts to the Drude case, when $I_z$ diverged as $z_0^{-3}$ at very small heights.

Taking only the leading terms in expansion of $\sigma_{\rm hd}$ at large $q$, we get the following asymptotic behavior:
\begin{equation}
   I(z_0) \approx \frac{i \pi}{2} \frac{\sigma_\omega }{c} \frac{k_s^2}{k_0 z_0^2},
\end{equation}
where $\sigma_\omega = n e^2/m_e \omega$ and we have introduced the wave vector of sound waves supported by 2DES, $k_s = \sqrt{2}\omega/v_0$. This asymptotic, shown in Fig.~\ref{fig:Hydro} with dashed line, matches well the full expression for $I_z$ at small heights.

\subsection{Ballistic transport}
Another limiting case for spatially dispersive conductivity of 2DES is realised for very long free paths, both for carrier collisions with disorder and with each other, $\omega\tau_p \gg 1$ and $\omega\tau_{ee} \gg 1$. If the frequencies and wave vectors still lie in the classical domain, $\omega \ll \varepsilon_F/v_0$ and $q \ll k_F$, the {\it ballistic} conductivtiy can be found from kinetic equation with the following result~\cite{Stern1967}:
\begin{equation}
\label{Sigma_bal}
\sigma_{\rm bal}=\frac{n e^2/m_e}{\sqrt{(\omega+i/\tau_p)^2-v_0^2 q^2}}
\end{equation}
For short wavelengths, $q > \omega/v_0$, the conductivity is purely real even at ultimately scarce collisions. This is the manifestation of Landau damping effect. In short-wavelength limit, the decay of ballistic conductivity ($\sigma_{\rm bal} \propto q^{-1}$) is intermediate between those in hydrodynamic ($\sigma_{\rm hd} \propto q^{-2}$) and Drude ($\sigma_D \propto q^{0}$) regimes. Taking such a limit for conductivity, we evaluate the asymptotics of $I_z$ at very small height:

\begin{figure}[ht]
\center{\includegraphics[width=0.9\linewidth]{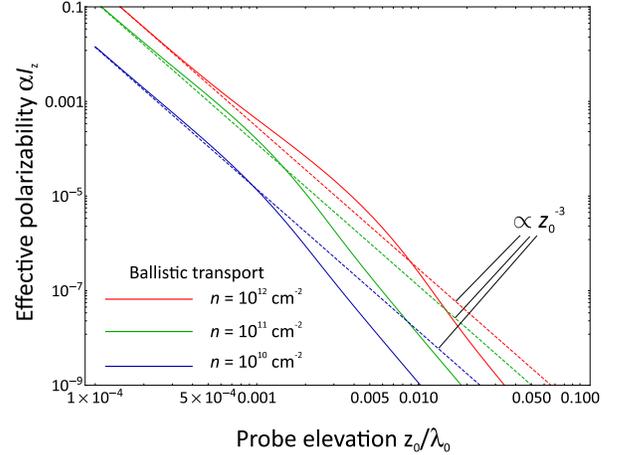}}
\caption{Effective polarizability of the near-field probe $\alpha I_z$ vs the probe elevation $z_0$ for ballistic transport in 2DES. Solid lines are results of exact integration (Es.~\ref{I}), dashed lines are power-law asymptotes. Incident wave frequency $\omega/2\pi = 2$ THz, momentum relaxation time $\tau_p = 2$ ps, effective mass $m = 0.067m_0$, background dielectric constant $\epsilon_b = 4$, tip radius $r_{\rm tip} = 10$ nm, Fermi velocity $v_0 = 10^6$ m/s}
\label{fig:IZbforn}
\end{figure}
\begin{equation}
    I_z \approx \frac{\varepsilon_b}{4z_0^3} \frac{p}{1 + i p}, \qquad p = \varepsilon_b^{-1} \frac{2\pi \sigma_\omega}{v_0}.
\end{equation}
At a first glance, the scaling of induced dipole moment $I_z \propto z_0^{-3}$ for ballistic transport is indifferent from that for Drude conductivity. However, ballistic conduction leads to a carrier density-dependent prefactor in induced dipole moment, while for Drude conduction the asymptotics is density-independent. Experimentally, these two transport modes can be conveniently distinguished by varying the 2D electron density with gate voltage.

\section{Discussion and conclusions}

We have demonstrated the possibility to distinguish between transport regimes in two-dimensional electron systems via the height dependence of near-field tip dipole moment. The height dependence follows the $z_0^{-3}$ law for classical ballistic and Drude regimes of conduction in 2DES. In the former case, the prefactor of power-law is dependent on the carrier density, in the latter case it is density-independent. For hydrodynamic regime of carrier transport, the build-up of dipole moment at small heights is not as rapid, and obeys the $z_0^{-3}$ law. 

The obtained results are valid both for graphene and 2DES with parabolic bands, such as quantum wells based on III-V compounds. In the former case, the carrier effective mass in expressions for the conductivity [(\ref{Sigma_drude}), (\ref{Sigma_hd}) and (\ref{Sigma_bal})] should be interpreted as $\varepsilon_F/v_0^2$, where $\varepsilon_F$ is the carrier Fermi energy and $v_0 = 10^6$ m/s is the constant Fermi velocity. In case of parabolic-band 2DES, the Fermi velocity is dependent on Fermi energy $v_0 = \sqrt{2 m_e \varepsilon_F}$, where $m_e$ is the constant effective mass.

It is possible to extend the discussion toward the quantum regimes of electron conductivity realized at $q \sim k_F$, where $k_F$ is the Fermi wave vector. While the detailed analysis of such case is beyond the scope of present communication, we note a very different scaling of ultra-quantum conductivities for graphene and parabolic-band 2DES. In the case of graphene, $\sigma(q \gg k_F) \propto q^{-1}$~\cite{DasSarma_plasmons}, while in the case of parabolic-band 2DES, $\sigma(q \gg k_F) \propto q^{-5}$~\cite{Stern1967}. A rapid drop in conductivity at large $q$ in parabolic-band 2DES should lead to height-independent behaviour of $I_z$ at $z_0 k_F \sim 1$.

\begin{figure}[ht]
\center{\includegraphics[width=0.9\linewidth]{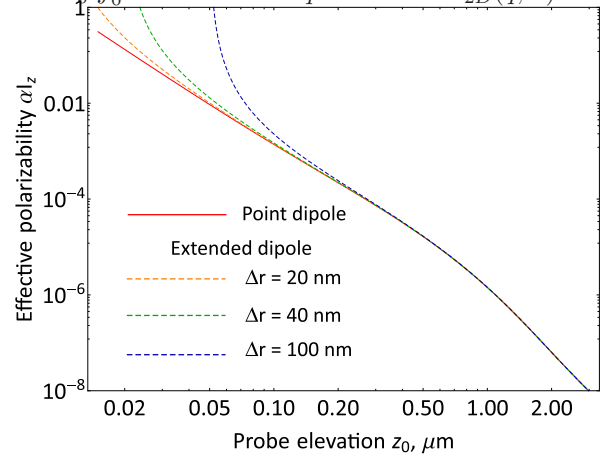}}
\caption{Effect of finite probe size on the scaling of effective polarizability $\alpha I_z$. Solid line corresponds to the point dipole model, dashed lines -- to the elongated dipole model with various sizes of the dipole $\Delta r$. All calculations correspond to the Drude conductivity with $n = 10^{12}$ cm$^{-2}$, $\omega/2\pi = 2$ THz, $\tau_p = 2$ ps, $\alpha = 10^{-24}$ cm$^{-3}$}
\label{fig:elongated}
\end{figure}

The presented calculation was performed for a point-dipole model of a near-field tip. We realize that interpretation of current experimental results requires more complex models, such as elongated dipole~\cite{Elongated_dipole}, exact conformal mappings~\cite{Monopole_SNOM}, or fully numerical simulations~\cite{Fogler_quant_near_field}. Without going into details of such models, we can account for 'elongation' of dipole at the probe tip by modelling it as two charges, $Q$ and $-Q$, separated by a finite distance $\Delta r$. The dipole moment can be still estimated as $Q \Delta r = \alpha E(z_0)$, where $\alpha$ is the tip polarizability and $E(z_0)$ is the electric field between the two charges. The expression for dipole moment in such situation is slightly modified:

\begin{equation}
\label{I_elong}
   I_z =\frac{1}{k_0} \int_0^{\infty}  \frac{ 2\pi i \sigma(q,\omega)}{c}\frac{\sinh q\Delta r}{q\Delta r} e^{-2 q z_0 } \frac{q^3dq}{ \epsilon_{2D}(q,\omega)},
\end{equation}

The resulting dependence of polarizability $I_z$ on height $z_0$ at various dipole elongations $\Delta r$ is shown in Fig.~\ref{fig:elongated}. Naturally, the range of accessible heights for such model is limited to $z_0 > \Delta r/2$. Otherwise, the lower charge falls below the 2DES plane, and the expression (\ref{I_elong}) formally diverges. It is also instructive that for small heights and relatively large elongation ($|z_0 - \Delta r/2| \ll z_0$) we restore the monopole model of the probe. In this case, only the lower charge interacts efficiently with the 2DES, and scaling of $I_z$ with height can be different.

\section{Acknowledgements}
The research was supported by the grant 21-72-10163 of the Russian Science Foundation.
The authors thank Alexey Y. Nikitin for helpful discussions.

\bibliography{dialogue}

\end{document}